\documentclass[letterpaper,notoc]{CECS}
\usepackage{epsfig}
\title{Black holes and asymptotics of 2+1 gravity
coupled to a scalar field}
\author{Marc Henneaux$^{1,2}$,  Cristi\'an Mart\'{\i}nez$^{1}$, Ricardo Troncoso$^{1}$ and Jorge Zanelli$^{1}$\footnote{{\it E-mail:} {\tt henneaux@ulb.ac.be, martinez@cecs.cl, ratron@cecs.cl, jz@cecs.cl}}\\ $^{1}$Centro de Estudios Cient\'{\i}ficos (CECS), Casilla 1469, Valdivia, Chile
\\$^{2}$Facult\'{e} des Sciences, Universit\'{e} Libre de
Bruxelles, Campus Plaine C.P.231, B-1050 Bruxelles, Belgium.  }

\preprint{{\tiny CECS-PHY-01/06} \\[-2.5mm] {\tiny ULB-TH-02/07}}

\abstract{
We consider 2+1 gravity minimally coupled to a self-interacting scalar field. The case in which
the fall-off of the fields at infinity is slower than that of a localized distribution of matter
is analyzed. It is found
that the asymptotic symmetry group remains the same as in pure gravity (i.e., the conformal group). The generators of the asymptotic symmetries, however, acquire a
contribution from the scalar field, but the algebra of the canonical generators possesses the
standard central extension. In this context, new massive black hole solutions with a regular
scalar field are found for a one-parameter family of potentials. These black holes are
continuously connected to the standard zero mass black hole. }

\begin{document}

\section{Introduction}

Three-dimensional Einstein theory with a negative cosmological constant has
 been the source of a
number of insights in gravitation physics, including the existence of black holes
\cite{BTZ,BHTZ}, their thermodynamics \cite{Carlip-Teitelboim,Strominger}, and the
AdS/CFT correspondence \cite{MAGOO}.

For matter-free gravity, the behavior of the three-dimensional metric at spatial infinity is given
by \cite{Brown-Henneaux}
\begin{equation}
\begin{array}{lllllll}
g_{rr} & = & \displaystyle \frac{l^{2}}{r^{2}}+O(r^{-4})\; &  & g_{tt} & = & \displaystyle
-\frac{r^{2} }{l^{2}}+O(1)\;
\\[2mm]  g_{tr} & = & O(r^{-3})\; &  & g_{\varphi \varphi } & = & r^{2}+O(1)\; \\[1mm] g_{\varphi r} & = &
O(r^{-3})\; &  & g_{t\varphi } & = & O(1)\;
\end{array}
\label{BHBC}
\end{equation}
where the cosmological constant is $\Lambda =-l^{-2}$. These conditions, in fact, hold not only
in the absence of matter but also for localized matter fields which fall off sufficiently fast at
infinity, so as to give no contributions to the surface integrals defining the generators of the
asymptotic symmetries.

The symmetry group which leaves invariant conditions (\ref{BHBC}) is generated by the following
asymptotic Killing vectors
\begin{eqnarray}
\eta ^{t} &=&l\left[ T^{+}+T^{-}+\frac{l^{2}}{2r^{2}}(\partial _{+}^{2}T^{+}+\partial
_{-}^{2}T^{-})\right] +O(r^{-4})\;,  \nonumber \\ \eta ^{r} &=&-r(\partial _{+}T^{+}+\partial
_{-}T^{-})+O(r^{-1})\;, \label{AsymptKillingVectors} \\ \eta ^{\varphi }
&=&T^{+}-T^{-}-\frac{l^{2}}{2r^{2}}(\partial _{+}^{2}T^{+}-\partial
_{-}^{2}T^{-})+O(r^{-4})\;,\nonumber
\end{eqnarray}
where $T^{+}(x^{+})$ and $T^{-}(x^{-})$ generate two independent copies of the Virasoro algebra and
$x^{\pm }=t/l\pm \varphi $. The charges that generate the asymptotic symmetries involve only the
metric and its derivatives, and are given by
\begin{equation}
Q(\xi )=\frac{1}{16\pi G}\int d\varphi \left\{ \xi ^{\perp }\left[ \frac{l}{r
}+\frac{r^{3}}{l^{3}}\left( g_{rr}-\frac{l^{2}}{r^{2}}\right) \right]   +
 \frac{ 1}{lr}\left( \xi ^{\perp }+r\xi ^{\perp },_{r}\right) \left[ g_{\varphi \varphi
}-r^{2}\right] +2\xi ^{\varphi }\pi _{\varphi }^{r}\right\} .
\end{equation}

The Poisson brackets algebra of the charges yields two copies of the Virasoro algebra with a
central charge \cite{Brown-Henneaux}
\begin{equation}
c=\frac{3l}{2G}\,.  \label{Central Charge}
\end{equation}

This fact provides a possible explanation for the microscopic origin of entropy of a spinning
black hole in $2+1$ dimensions \cite{Strominger}.

\section{Matter, asymptotics and global charges}

There are instances in which the matter fields modify the asymptotic behavior of the
metric, as it occurs for the electrically charged black hole, in which case the metric has a
logarithmic divergence\cite{BTZ,MTZ}. This brings about the potential danger of
having divergent contributions coming both from the gravitational and matter actions. In that
case, the asymptotic conditions must be such that the sum of both contributions converges. Here we
analyze this possibility in a simplified setting, where matter is  given by a single
self-interacting scalar field minimally coupled to three dimensional gravity, with the action
\begin{equation}
I[g,\phi ]=\frac{1}{\pi G}\int d^{3}x\sqrt{-g}\left[ \frac{R}{16}-\frac{1}{2}
(\nabla \phi )^{2}-V(\phi )\right] \;.  \label{Action}
\end{equation}

The standard three-dimensional black hole is a solution of this action with a constant vanishing
field $\phi =0$  with $V(0)=\Lambda /8$ and $V'(0)=0$. With the asymptotic 
conditions (\ref{BHBC}), the scalar
field would not contribute to the charges if for large $r$ it decays at least as $\phi \sim
r^{-(1+\varepsilon )}${\bf .} There exist, however, black hole solutions with a scalar field which
is regular everywhere and behaves like $\phi \sim r^{-1/2}$ for $r\rightarrow \infty $ (see
below), and therefore the charges must be computed from scratch. The asymptotic behavior of these
solutions belongs to the following class
\begin{equation}
\phi =\frac{\chi }{r^{1/2}}-\frac{2}{3}\frac{\chi ^{3}}{r^{3/2}}
+O(r^{-5/2})\label{AsymptPhi}
\end{equation}
\begin{equation}
\begin{array}{lll}
g_{rr}= \displaystyle \frac{l^{2}}{r^{2}}-\frac{4l^{2}\chi ^{2}}{r^{3}}+O(r^{-4}) & &
\displaystyle g_{tt} = -\frac{r^{2}}{l^{2}}+O(1)  \\[2mm] g_{tr} = O(r^{-2}) &  & g_{\varphi
\varphi } =  r^{2}+O(1) \\[1mm] g_{\varphi r} = O(r^{-2}) &  & g_{t\varphi } = O(1)
\end{array}
\label{AsymptGeneral}
\end{equation}
where $\chi =\chi (t,\varphi )$, and we have set $G=1$. The asymptotic behavior of $g_{r\mu }$ has
a slower falloff than in Eq. (\ref{BHBC}). Remarkably, this set of conditions is also left
invariant under the Virasoro algebra generated by the asymptotic Killing vectors (\ref
{AsymptKillingVectors}).

The contributions of gravity and the scalar field to the conserved charges, $
Q_{G}(\xi )$ and $Q_{\phi }(\xi )$, can be found using the Regge-Teitelboim approach
\cite{Regge-Teitelboim}. Making use of the asymptotic conditions, their variations can be shown to
be of the form \cite{Footnote1}

\begin{eqnarray}
\delta Q_{G}(\xi ) &=&\frac{1}{16\pi }\int d\varphi \left\{ \xi ^{\bot }\left[ \frac{2}{lr}\delta
g_{\varphi \varphi }+rg^{-1/2}g^{rr}\delta g_{rr}\right]  +2\xi
^{\varphi }\delta g_{\varphi t}\right\}  \nonumber \\ &=&\frac{1}{16\pi }\int d\varphi \left\{
\frac{1}{lr}\xi ^{\bot }\delta g_{\varphi \varphi }-2r\xi ^{\bot }\delta (g^{-1/2}) 
+2\xi ^{\varphi }\delta g_{\varphi t}\right\}\label{DeltaQG} \;.
\end{eqnarray}
\begin{eqnarray}
\delta Q_{\phi }(\xi ) &=&-\frac{1}{\pi }\int d\varphi \left[ \xi ^{\bot }g^{1/2}g^{rr}\partial
_{r}\phi \delta \phi \right]  \nonumber  \\
&=&\frac{1}{4\pi l}\int d\varphi \;\xi ^{\bot }\left[ \delta \chi ^{2}-\frac{
1}{r}\delta \chi ^{4}\right]\label{DeltaQPhi} \;.
\end{eqnarray}
Asymptotically, $\xi ^{\bot }\sim \xi ^{t}r/l$ grows linearly with $r$ and hence the first term in
$\delta Q_{\phi }$ in Eq. (\ref{DeltaQPhi}) is linearly divergent. This term cancels a divergent
piece in $\delta Q_{G}$ coming from $\delta (g^{-1/2})$ because, to leading order,  $\delta
g_{rr}=-4l^{2}r^{-3}\delta \chi ^{2}$. In other words, if $g_{rr}$ had been assumed independent of
$\chi $, this cancellation would not have occurred.

Hence, the total charge $Q=Q_{G}+Q_{\phi }$ can be integrated to obtain
\begin{equation}
Q(\xi )=\frac{1}{16\pi }\int d\varphi \left\{ \frac{\xi ^{\bot }}{lr}\left(
(g_{\varphi \varphi }-r^{2})-2r^{2}(lg^{-1/2}-1)  +4r^{2}\left[ \phi ^{2}+
\frac{1}{3}\phi ^{4}\right] \right) +2\xi ^{\varphi }\pi _{\varphi }^{r}\right\} \,,
\label{Qtotal}
\end{equation}
where the background configuration ($Q(\xi )=0$) has been chosen to be the standard massless $2+1$
black hole with vanishing scalar field. For the class of potentials which are consistent with the
modified asymptotic behavior (\ref{AsymptGeneral}), this configuration corresponds to the ground
state. A similar expression for $Q(\xi )$ could also be found using covariant methods as, for
instance, in \cite{Glenn}.

It should be emphasized that the algebra of the charges (\ref{Qtotal}) is identical to the one
found in \cite{Brown-Henneaux}, namely, two copies of the Virasoro algebra with a central
extension. Indeed, it follows from the work in Ref \cite{Brown-Henneaux2}, that the
bracket of two charges provides a realization of the asymptotic symmetry algebra with a possible
central extension. Thus, one only needs to determine the central charge, which is done by
computing the variation of the charges on the vacuum. Here the vacuum is the same as in pure
gravity, and therefore the computation is identical,
\begin{eqnarray*}
\delta _{\eta }Q[\tilde{\eta}] &=&\frac{1}{8\pi }\int d\varphi \left\{ \frac{
1}{l^{2}}\tilde{\eta}^{t}\delta _{\eta }g_{\varphi \varphi }+\tilde{\eta}
^{\varphi }\delta _{\eta }g_{\varphi t}\right\} \; \\
&=&\frac{-l}{8\pi }\int d\varphi \left[ \tilde{T}^{+}\partial _{+}^{3}T^{+}+
\tilde{T}^{-}\partial _{-}^{3}T^{-}\right] \,,
\end{eqnarray*}
which shows that the value of the central charge remains unchanged.

\section{Exact black hole solutions}

An exact solution for which the metric and the scalar field satisfy asymptotic conditions
(\ref{AsymptGeneral}) is obtained for a particular one-parameter family of potentials of the form
\begin{equation}
V_{\nu }(\phi )=-\frac{1}{8l^{2}}\left( \cosh ^{6}\phi +\nu \sinh ^{6}\phi \right) \;.  \label{The
Potential}
\end{equation}
This expression for{\bf \ }$V_{\nu }(\phi )$ is obtained through the scale transformation
$\hat{g}_{\mu \nu }=\Omega ^{-2}g_{\mu \nu }$ in the action
\begin{equation}
I[\hat{g},\hat{\phi}]=\frac{1}{\pi }\int d^{3}x\sqrt{-\hat{g}}\left( \frac{
\hat{R}+2l^{-2}}{16}  - \frac{1}{2}(\nabla \hat{\phi})^{2}-\frac{1}{16}\hat{R}
\hat{\phi}^{2}+\frac{\nu }{8l^{2}}\hat{\phi}^{6}\right) \;, \label{Conformal Action}
\end{equation}
for $\Omega =(1-\hat{\phi}^{2})$, where the scalar field has been redefined as $\hat{\phi}=\tanh
\phi .$\ In this frame, the matter piece of the action is conformally invariant -- i.e., it
is unchanged under $\hat{g}_{\mu \nu }\rightarrow \lambda ^{2}(x)\hat{g}_{\mu \nu }$ and
$\hat{\phi}\rightarrow \lambda ^{-1/2}\hat{\phi}$ --, for any value of $\nu $.

For $\nu \geq -1$, there is a solution that describes a static and circularly symmetric
black hole, dressed with a scalar field which is regular everywhere, given by

\begin{equation}
\phi ={\rm arctanh}\sqrt{\frac{B}{H(r)+B}}\;,  \label{Scalar}
\end{equation}
where $B$ is a non-negative integration constant and
\[
H(r)=\frac{1}{2}\left( r+\sqrt{r^{2}+4Br}\right) \;.
\]
The line element reads
\begin{equation}
ds^{2}=-\left( \frac{H}{H+B}\right) ^{2}F(r)dt^{2}+\left( \frac{H+B}{H+2B}
\right) ^{2}\frac{dr^{2}}{F(r)}+r^{2}d\varphi ^{2}\;,  \label{The metric}
\end{equation}
with
\[
F=\frac{H^{2}}{l^{2}}-(1+\nu )\left( \frac{3B^{2}}{l^{2}}+\frac{2B^{3}}{
l^{2}H}\right) \;.
\]
The causal structure of this geometry is the same as for the standard, nonrotating $2+1$ black
hole. The event horizon is located at

\[
r_{+}=B\Theta _{\nu }\,,
\]
where the constant $\Theta _{\nu }${\bf \ }is the first zero of the Schuster function of order
$\nu $, given by \cite{Schuster}\
\[
\Theta _{\nu }=2(z\bar{z})^{2/3}\frac{z^{2/3}-\bar{z}^{2/3}}{z-\bar{z}}\;,
\]
and $z=1+i\sqrt{\nu }$. As a function of $\nu $, $\Theta _{\nu }$ is monotonically increasing, and
asymptotically grows as $\sqrt{\nu }$ (see Fig. 1).

\begin{figure}[tbm]
\begin{center}
  \leavevmode
  \epsfxsize=3 in
\epsfbox{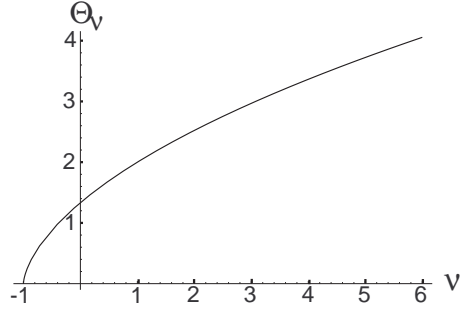}\caption{Plot of  $\Theta _{\nu }$ as a function of $\nu $. The intercepts are
at $\Theta_{-1}=0$ and $\Theta_{0}=\frac{4}{3}$. }
\end{center}
\end{figure}

The Hawking temperature for this black hole is
\begin{equation}
T=\frac{3B}{2\pi l^{2}}\frac{(1+\nu )}{\Theta _{\nu }}\;, \label{Temperature}
\end{equation}
and the mass can be obtained from Eq. (\ref{Qtotal}) for $\xi =\partial _{t}$ ,
\begin{equation}
M=Q(\partial _{t})=\frac{3B^{2}}{8l^{2}}(1+\nu )\;.  \label{Mass}
\end{equation}
The entropy is obtained evaluating the Euclidean action at the saddle point or, equivalently, from
the first law of thermodynamics,
\begin{equation}
S=\frac{\pi r_{+}}{2}=\frac{A}{4}\;.  \label{Entropy}
\end{equation}
Note that the specific heat is equal to the entropy $C=\partial M/\partial T=
\frac{\pi }{2}r_{+}$, exactly as for the standard three dimensional black hole. The positivity of
$C$ implies that the black hole can always reach the thermal equilibrium with a heat bath at some
fixed temperature.

The one parameter family of potentials (\ref{The Potential}) behaves as
\[
V_{\nu }\rightarrow -\frac{(1+\nu )}{2^{9}l^{2}}\exp (6\phi )\;,
\]
as $\phi >>1$ for $\nu >-1$. However, when $\nu =-1$ the potential behaves as
\[
V_{-1}\rightarrow -\frac{1}{l^{2}}\frac{3}{2^{7}}\exp (4\phi )\;,
\]
and the metric (\ref{The metric}) reads
\begin{equation}
ds^{2}=-\frac{r^{2}}{l^{2}}dt^{2}+\frac{l^{2}}{r^{2}+4Br}dr^{2}+r^{2}d%
\varphi ^{2}\;.  \label{Special case}
\end{equation}
This line element shares the same causal structure with the standard massless three dimensional
black hole. In fact, it is also left invariant under boosts in the $t-\varphi $ plane. However,
for the metric (\ref {Special case}) the integration constant $B$ is not related to the mass. In
fact, by virtue of Eqs. (\ref{Mass}), (\ref{Temperature}),  (\ref{Entropy}) the mass and the
temperature, as well as the entropy,  vanish for $\nu =-1$. Since $Q(\xi )$ in Eq. (\ref{Qtotal})
vanishes, this configuration can be regarded as a degenerate ground state.

In the conformal frame and for $\nu >-1$, the scalar field is
\begin{equation}
\hat{\phi}=\sqrt{\frac{B}{\rho +B}}\;,  \label{Phi-gorro}
\end{equation}
and the metric is given by $d\hat{s}^{2}=-Fdt^{2}+F^{-1}d\rho ^{2}+\rho ^{2}d\varphi ^{2}$,
with
\[
F=\frac{\rho ^{2}}{l^{2}}-(1+\nu )\left( \frac{3B^{2}}{l^{2}}+\frac{2B^{3}}{
l^{2}\rho }\right) \;,
\]
where the radial coordinate has been changed as $r=\rho ^{2}(\rho +B)^{-1}$. This metric describes
a black hole whose horizon radius is
\begin{equation}
\rho _{+}=\frac{B}{2}\left( \Theta _{\nu }+\sqrt{\Theta _{\nu }^{2}+4\Theta _{\nu }}\right) \;.
\end{equation}

In this frame, the entropy (\ref{Entropy}) reads
\[
\hat{S}=\frac{\hat{A}}{4}\cdot \frac{2}{1+\sqrt{1+4/\Theta _{\nu }}}\leq \frac{\hat{A}}{4}\,.
\]

For $\nu =-1$ the metric is the standard zero-mass $2+1$ black hole, and the scalar field does
not contribute to the stress-energy tensor ($T_{\mu \nu }=0$). For $\nu =0$\ the solution reduces
to the one found \cite{M-Z} in the conformal frame ($\hat{g},\hat{\phi}$) and the central charge
in this case was computed in Ref. \cite{Natsuume:1999cd}.

\section{Discussion}
~~~~\hskip 2mm
{\bf 1.} It should be noted that the scalar field cannot be switched off
keeping the mass fixed. In fact there is only one integration constant ($B$
), and for $\phi \rightarrow 0$ , the geometry approaches the massless black hole. On the other
hand, for a given mass there are two widely different black hole solutions, one with $\phi =0$
(the standard 2+1 black hole) and one with a nontrivial scalar field given by Eq. (\ref{Scalar}) in
the Einstein frame, or Eq. (\ref{Phi-gorro}) in the conformal frame.

{\bf 2. }The black hole solution discussed here is similar to the black hole solution coupled to a
conformal scalar field in asymptotically flat 3+1 spacetime \cite{Bekenstein}. In both cases there
is only one integration constant which simultaneously fixes the mass and the value of the scalar
field. However the geometry in \cite{Bekenstein} is that of an extremal black hole, and the scalar
field is softly divergent at the horizon.

{\bf 3.} The fact that the algebra of the charges has the same central extension as in the standard
case would suggest that the number of degenerate microstates ${\cal N}(L_{0},\bar{L}_{0},c)$
could be given by Cardy's formula \cite{Strominger}. However, if one naively follows this
approach, one finds
\[
S=\log {\cal N}=2\pi l\sqrt{\frac{M}{2}}=\frac{\sqrt{3(1+\nu )}}{\Theta _{\nu }}\frac{A}{4}\,.
\]
However, since $\sqrt{3(1+\nu )}/\Theta _{\nu }>1$ for $\nu >-1$ this computation would imply
$S>\frac{A}{4}$, in contradiction with Hawking's result (the case $\nu =-1$ is{\bf \ }empty since
for $\nu =-1$ the area vanishes).

{\bf 4.} The asymptotic expansion for the scalar field in Eq. (\ref {AsymptPhi}) can be relaxed as
\begin{equation}
\phi =\frac{\chi }{r^{1/2}}+\frac{\beta (\chi )}{r^{3/2}}+O(r^{-5/2})\;,  \label{beta}
\end{equation}
however, invariance of the asymptotic conditions under the Virasoro symmetry implies $\beta (\chi
)=\alpha \chi ^{3}$, where $\alpha $ is an arbitrary real number without variation. The solutions
corresponding to the class of potentials $V_{\nu }(\phi )$\ considered here have $\alpha =-2/3$.

Changing the asymptotic conditions according to Eq. (\ref{beta}) modifies the expression for the term $\phi ^{4}$\ in the charge as
\begin{equation}
Q(\xi )=\frac{1}{16\pi }\int d\varphi \left\{ \frac{\xi ^{\bot }}{lr}\left( (g_{\varphi \varphi
}-r^{2})-2r^{2}(lg^{-1/2}-1)   +4r^{2}\left[ \phi
^{2}+(1+\alpha )\phi ^{4}\right] \right) +2\xi ^{\varphi }\pi _{\varphi }^{r}\right\} \,,
\end{equation}
which is finite and leads to the same central charge independently of $\alpha $.

{\bf 5.} Consistency of the asymptotic boundary conditions introduced here with the field equations
is sufficient to fix the potential $V(\phi )$ to be of the form
\begin{equation}
V(\phi )=-\frac{1}{8l^{2}}-\frac{3}{8l^{2}}\phi ^{2}-\frac{1}{2l^{2}}\phi ^{4}+\phi ^{6}U(\phi
)\;,  \label{Generic Potential}
\end{equation}
where $U(\phi ^{2})$\ could be any smooth function around $\phi =0$. In spite of the fact
that $V(\phi )$\ could even be unbounded from below, this potential satisfies the stability bound
that guarantees the
perturbative stability of AdS space\cite{B-F,M-T}. The potential $
V_{\nu }(\phi )$ in Eq. (\ref{The Potential}) belongs to this family, and for different values of
the dimensionless parameter $\nu $,  different forms of $U(\phi ^{2})$ are obtained. The exact
connection between $V(\phi )$\ and $\alpha $\ is an open question.

\acknowledgments

We thank Professor Claudio Teitelboim for useful discussions and enlightening comments. This
research is partially funded by FONDECYT grants 1020629, 1010446, 1010449, 1010450, 7010446, and 7010450,  and from the generous support to CECS by Dimacofi and Empresas CMPC. 
The work of M. H.  is partially
supported by the ``Actions de Recherche Concert{\'{e}}es'' of the ``Direction de la Recherche
Scientifique - Communaut{\'{e}} Fran{\c{c}}aise de Belgique'', by IISN - Belgium (convention
4.4505.86). CECS is a Millennium Science Institute.


\begin{thebibliography}{99}
\bibitem{BTZ}  M.~Ba\~{n}ados, C.~Teitelboim and J.~Zanelli, Phys.\ Rev.\
Lett.\ {\bf 69}, 1849 (1992).

\bibitem{BHTZ}  M.~Ba\~{n}ados, M.~Henneaux, C.~Teitelboim and J.~Zanelli,
Phys.\ Rev.\ {\bf D48}, 1506 (1993).

\bibitem{Carlip-Teitelboim}  S. Carlip and C. Teitelboim, Phys.Rev.{\bf D51}
, 622 (1995).

\bibitem{Strominger}  A. Strominger, JHEP {\bf 9802}, 009 (1998).

\bibitem{MAGOO}  For a recent review, see O. Aharony, S. Gubser, J.
Maldacena, H. Ooguri, Y. Oz, Phys.Rept. {\bf 323} (2000),183-386.

\bibitem{Brown-Henneaux}  J.~D.~Brown and M.~Henneaux, Commun.\ Math.\
Phys.\ {\bf 104}, 207 (1986).

\bibitem{MTZ}  C. Mart\'{\i }nez, C. Teitelboim and J. Zanelli, Phys. Rev.
{\bf D61}, 104013 (2000).

\bibitem{Regge-Teitelboim}  T. Regge and C. Teitelboim, Ann. Phys. (N.Y.)
{\bf 88 }(1974) 286.

\bibitem{Footnote1}  The asymptotic behavior of the momenta is given by $\pi
^{rr}=O(1)$, $\pi ^{r\varphi }=O(r^{-2})$, $\pi ^{\varphi \varphi }=O(r^{-4}) $, and $\pi _{\phi
}=O(r^{-3/2})$. Note that $\pi _{\varphi }^{r}=g_{\varphi t}$ to leading order.

\bibitem{Glenn}  G. Barnich and F. Brandt, {\it Covariant theory of
asymptotic symmetries, conservation laws and central charges,} hep-th/0111246.

\bibitem{Brown-Henneaux2}  J.D. Brown and M. Henneaux, J.Math.Phys{\bf . 27 }
(1986){\bf \ }489.

\bibitem{Schuster}  E. Schuster, Acta Phil. Vald. {\bf 19} (1877) 31.

\bibitem{M-Z}  C.~Mart\'{\i }nez and J.~Zanelli, Phys.\ Rev.\ D {\bf 54},
3830 (1996).

\bibitem{Natsuume:1999cd}  M.~Natsuume, T.~Okamura and M.~Sato, Phys.\ Rev.\
D {\bf 61}, 104005 (2000).

\bibitem{Bekenstein}  N. Bocharova, K. Bronikov and V. Melnikov, Vestn.
Mosk. Univ. Fiz. Astron. {\bf 6} (1970) 706. J. D. Bekenstein, Ann. Phys. (N. Y.){\bf \ 82} (1974)
535; Ann. Phys. (N. Y.) {\bf 91} (1975) 72.

\bibitem{B-F}  P. Breitenlohner and D. Z. Freedman, Phys. Lett. B {\bf 115}
(1982) 197; Ann. Phys. (N.Y.) {\bf 144} (1982) 249.

\bibitem{M-T}  L. Mezincescu and P. K. Townsend, Ann. Phys. (N.Y.) {\bf 160 }
(1985) 406.
\end{thebibliography}
\end{document}